\documentclass[%
 reprint,
 amsmath,amssymb,
 aps,
]{revtex4-2}

\usepackage{graphicx}
\usepackage{dcolumn}
\usepackage{bm}

\begin{document}


\title{Topological metal state in helimagnets}

\author{Yu.\ B.\ Kudasov}

\affiliation{Sarov Physics and Technology Institute NRNU MEPhI, 6 Dukhov str., Sarov, 607186, Russia}
\affiliation{Russian Federal Nuclear Center - VNIIEF, 37 Mira pr., Sarov, 607188, Russia}%

\date{\today}

\begin{abstract}
A theory for nontrivial topology of band structure in metallic helimagnets is developed. Two theorems on electron dispersion in helimagnets are proved. They reveal a Kramers-like degeneracy in helical magnetic field. The generalized Bloch theorem together with periodic boundary conditions lead to a nontrivial topological band structure. As a result, an unusual spin structure of electron bands appears. A 2D model of nearly free electrons is proposed to describe conductive hexagonal palladium layers under an effective field of magnetically ordered CrO$_2$ spacers in PdCrO$_2$. The spin texture of the Fermi surface induces abnormal conductivity and nonreciprocity of electronic transport. It is shown that the topological metal state is closely related to unconventional antiferromagnets.
\end{abstract}

\maketitle
\section{Introduction}

The topology of band structure of crystalline solids was intensively studied during the last decades \cite{Hasan,Bansil}. An elegant technique was developed to transform the phase of Bloch wave function to a vector field called the Berry curvature \cite{Berry}. Due to periodic boundary conditions, the Brillouin zone can be represented as a circle (for a 1D chain), torus (for a 2D structure), or hypertorus (for 3D one). That is why, a distribution of the Berry curvature over the Brillouin zone may belong to distinct topological classes. Those of them, which differ from the vacuum state, are call nontrivial. Presently, all the topological classes are thoroughly classified and investigated \cite{Chiu}. Main theoretical and experimental efforts were concentrated on topological insulators and their edge states \cite{Tkachev,Hasan} because in this case topological features give rise to observable phenomena like surface states with extremely high mobility of charge carriers. Another problem that was thoroughly studied was a Kramers-type degeneracy in various magnetic and topological systems \cite{Ramazashvili}. Recently, magnetic topological insulators and semimetals including helimagnets turned out to be the focus of attention \cite{Bernevig,Yao}.

The present article addresses mainly to the following problem. Are nontrivial topological band structures possible that are not related to the Berry phase and Berry connection? Below we give a positive answer to this question, provide an example of such a system in helimagnets and discuss observable phenomena that arise from the novel topological structures.

A motion of spin-1/2 particle in helical magnetic field was studied for a long time \cite{Dzyaloshinskii,Calvo,Calvo2,Fresard,Fraerman,Kishine,Kudasov }. In particular, an exact solution is known \cite{Calvo} and there are various approximate approaches \cite{Calvo2,Kudasov}. It was proved that dispersion curves (surfaces) in helimagnets have a specific symmetry \cite{KudasovPRB} related to the time reversal. The electron transport in noncolinear magnetic structures exhibits nonreciprocity due to breaking of spatial inversion symmetry \cite{Akaike,Fraerman}. It has long been known that a gapless band structure is formed under helical magnetic field \cite{Dzyaloshinskii,Calvo}, but only recently it was shown that this is a consequence of nontrivial band topology \cite{KudasovJETPL2}. 

A renewed interest in electron dispersion of complex magnetic systems was inspired by the recent discovery of uncorrelated nonrelativistic antiferromagnetic phases with strong spin splitting at zero net magnetization \cite{Smejkal,Hellenes,Brekke}. Among them there are colinear  (altermagnets) \cite{Smejkal} and other unconventional structures  ($p$-wave antiferromagnets and etc.) \cite{Hellenes,Brekke}. 

A spin space group (SSG) theory is a useful instrument for investigation of helical magnetic systems \cite{Brinkman,Sandratskii,Liu}. A SSG operator $\{\mathbf{ \boldsymbol\alpha}|\mathbf{ \boldsymbol\beta}|\mathbf{t}\}$ comprises the spin rotation $\mathbf{ \boldsymbol\alpha}$ and space transformation $\{\mathbf{ \boldsymbol\beta}|\mathbf{t}\}$ which combines the rotation $\mathbf{ \boldsymbol\beta}$ and translation $\mathbf{t}$. In the framework of the theory, generalized translations are introduced ($\{\mathbf{ \boldsymbol\alpha}|0|\mathbf{t}\}$), and a generalized Bloch theorem was proved \cite{Brinkman,Sandratskii}.The SSG theory occurs especially useful in case of nonrelativistic studies. 

Below, PdCrO$_2$ is assumed to be a model substance where the topological structure under consideration is realized. It belongs to metallic delafossites  (PtCoO$_2$, PdCoO$_2$, PdCrO$_2$), which are layered compounds with anomalous transport properties \cite{Mackenzie,Akaike,Takatsu,Hicks}. Their in-plane conductivity at room temperature approaches to that of the best elementary conductors such as silver, copper, and aluminum \cite{Mackenzie}. CoO$_2$ and CrO$_2$ are proved to be dielectric interlayers \cite{Lechermann}, and the conductivity in these substances is determined entirely by hexagonal palladium or platinum layers. The combination of high conductivity and low mobile carrier density leads to extremely large values of electron mean free path, up to 20~m$\mu$ at low temperatures \cite{Hicks}. When electron momentum-relaxing scattering by
impurities or phonons is much weaker as compared to momentum-conserving scattering, a hydrodynamic regime of electron transport appears \cite{Scaffidi} as it was observed in PdCoO$_2$ \cite{Moll}.
The unusual behavior of the delafossites suggests a novel mechanism of electron transport \cite{Usui}.

A long-range magnetic order appears in PdCrO$_2$ below $T_c = 37.5$~K \cite{Mekata,Takatsu2,Billington}. Chromium ions form a 120$^0$ magnetic structure within a single layer. A complex interlayer arrangement forms 18 magnetic sublattices \cite{Takatsu2}. The appearance of the magnetic order is accompanied by a resistivity drop \cite{Daou}. Thus, a main hypothesis, which is going to be proved in the present article, is that the helical magnetic order induces the high-conductivity state under certain conditions. It should also be mentioned that unconventional anomalous Hall effect (UAHE) \cite{Takatsu3} and nonreciprocal electronic transport \cite{Akaike} are observed in the magnetically ordered state of PdCrO$_2$. Above $T_c$ a short-range magnetic order in hexagonal chromium layers persists up to about 500~K \cite{Daou,Billington}, and this is another surprising fact in view of the extremely high conductivity.    

The Fermi surface in the metallic delafossites was thoroughly investigated \cite{Hicks,Mackenzie}. In the paramagnetic state, it is quasi-two-dimensional with a single $\alpha$-orbit. The transition to the ordered state in PdCrO$_2$ leads to a reconstruction of the Fermi surface within the magnetic Brillouin zone and appearance of additional $\gamma$-orbits corresponding to pockets in the vicinity of K points \cite{Billington,Hicks2}.  

The theory developed in this article is based on two approaches presented recently in brief reports \cite{KudasovPRB,KudasovJETPL2}. The next two sections, which form a basis for theoretical description of topological band structure in helimagnets, rested upon them. Then, a simple 2D model for spin-textured Fermi surface is applied to investigate transport phenomena, especially nonreciprocal conductivity observed in PdCrO$_2$ \cite{Akaike}. The final section is devoted to a general discussion of novel topological band structure and possible ways of its experimental realization.

\section{Time-reversal symmetry in helimagnets}

\subsection{Two theorems on electron dispersion in helimagnets}

Let us consider a particle of spin-1/2 in a magnetic field, which is invariant under translation with period $\mathbf{T}$, i.e. $\mathbf{h}(\mathbf{r}+\mathbf{T})=\mathbf{h}(\mathbf{r})$. If we take into account an interaction between the spin and magnetic field, then the Hamiltonian has the form
\begin{eqnarray}
	\hat{H}=\hat{H}_0 + \mathbf{h}(\mathbf{r}) \hat{\mathbf{ \boldsymbol\sigma}} \label{H}
\end{eqnarray}
where $\hat{\mathbf{\boldsymbol\sigma}}$ are the Pauli matrices. One can prove two theorems on eigenvalues of this Hamiltonian \cite{KudasovPRB}.

{\bf{Theorem 1.}} If $\hat{H}_0$ is invariant under the operation $\hat{\mathbf{T}}_{1/2} \hat{\mathbf{\theta}}$, where $\hat{\mathbf{T}}_{1/2}$ is the space translation along vector $\mathbf{T}/2$, $\hat{\mathbf{\theta}}$ is the time-reversal operator, and $\mathbf{h}(\mathbf{r}+\mathbf{T}/2)=-\mathbf{h}(\mathbf{r})$, then eigenvalues $\varepsilon _ {\mathbf{k}}$ of $\hat{H}$ are at least two-fold degenerate for all the wave vectors $\mathbf{k}$ except those, which satisfy
\begin{eqnarray}
	\exp\big(i \mathbf{k} \mathbf{T}\big)=-1, \label{exp1}
\end{eqnarray}
and the following condition is fulfilled for all the eigenvalues: 
\begin{eqnarray}
	\varepsilon _ {\mathbf{k}, \langle \boldsymbol\sigma \rangle} = \varepsilon _ {-\mathbf{k},- \langle \boldsymbol\sigma \rangle} \label{T1}
\end{eqnarray}
where $ \langle \boldsymbol\sigma \rangle \equiv \langle \psi_ {\mathbf{k}} | \hat{ \boldsymbol\sigma} | \psi_ {\mathbf{k}} \rangle$ and $| \psi_ {\mathbf{k}} \rangle $ is the eigenstate of $\hat{H}$.

{\bf{Proof.}}  In the presence of magnetic field a symmetry operation containing the time reversal should be a product of $\hat{\mathbf{\theta}}$ and an operator, which change sign of the magnetic field \cite{Wigner}. Therefore, $\hat{\mathbf{T}}_{1/2} \hat{\mathbf{\theta}}$ is a symmetry operation for $\hat{H}$. Let us consider two eigenstates: $| \psi_ {\mathbf{k}} \rangle$ and $\hat{\mathbf{T}}_{1/2} \hat{\mathbf{\theta}} | \psi_ {\mathbf{k}} \rangle$. According to rules for antiunitary operators \cite{Messiah} we obtain
\begin{eqnarray}
	\langle \psi_ {\mathbf{k}} | \big( \hat{\mathbf{T}}_{1/2} \hat{\mathbf{\theta}} | \psi_ {\mathbf{k}} \rangle \big) =  \overline{\big(\langle \psi_ {\mathbf{k}} |\hat{\mathbf{\theta}}^\dagger \big) \big( \hat{\mathbf{T}}_{1/2} \hat{\mathbf{\theta}}^2 | \psi_ {\mathbf{k}} \rangle \big)} \nonumber \\ =- \langle \psi_ {\mathbf{k}} | \hat{\mathbf{T}}^\dagger  \big( \hat{\mathbf{T}}_{1/2} \hat{\mathbf{\theta}} | \psi_ {\mathbf{k}} \rangle \big).  \label{H11}
\end{eqnarray} 
By means of Bloch's theorem $ \hat{\mathbf{T}}| \psi_ {\mathbf{k}} \rangle = \exp\big(i \mathbf{k} \mathbf{T}\big)| \psi_ {\mathbf{k}} \rangle$ it is reduced to
\begin{eqnarray}
	\langle \psi_ {\mathbf{k}} | \big( \hat{\mathbf{T}}_{1/2} \hat{\mathbf{\theta}} | \psi_ {\mathbf{k}} \rangle \big) = -\exp(-i \mathbf{k} \mathbf{T}) \langle \psi_ {\mathbf{k}} | \big( \hat{\mathbf{T}}_{1/2} \hat{\mathbf{\theta}} | \psi_ {\mathbf{k}} \rangle \big).  \label{H12}
\end{eqnarray}
From here one can see that $| \psi_ {\mathbf{k}} \rangle$ and $\hat{\mathbf{T}}_{1/2} \hat{\mathbf{\theta}} | \psi_ {\mathbf{k}} \rangle$ are orthogonal and make up a pair of degenerate states for all $\mathbf{k}$ except those, which satisfy Eq.(\ref{exp1}). In the last case,
the state can be either degenerate or nondegenerate. Since operators $\hat{\mathbf{k}}$ and $\hat{\sigma}$ commute with $\hat{\mathbf{T}}_{1/2}$ and change sign under the time reversal, we obtain Eq.~(\ref{T1}) $\blacksquare$. 

The theorem can be extended to an interaction of the particle charge  $q_0$ and helical magnetic field by means of the substitution \begin{eqnarray}
	\hat{\mathbf{p}} \rightarrow \hat{\mathbf{p}} + q_0 \mathbf{A}(\mathbf{r}) \hat{\mathrm{I}}, \label{subst}
\end{eqnarray}
where $\mathbf{A}(\mathbf{r})$ is the vector potential, and $\hat{\mathrm{I}}$ is the unit matrix. 

Since the momentum operator $\hat{\mathbf{p}}$ is invariant with respect to translations and changes the sign under the time reversal, it enters Hamiltonian $\hat{H}_0$ to an even power or as a product $\hat{\mathbf{p}}$ and an operator $ \hat{f}$, which meets the following condition: $ \hat{f} = - \hat{\mathbf{T}}_{1/2} \hat{\mathbf{\theta}} \hat{f} \hat{\mathbf{\theta}}^{-1} \hat{\mathbf{T}}_{1/2}^{-1} $. That is why, if the vector potential corresponding to $\mathbf{h}(\mathbf{r})$ has the same translational symmetry under a proper gauge, i.e. 
\begin{eqnarray}
	\mathbf{A}(\mathbf{r}+\mathbf{T/2})=-\mathbf{A}(\mathbf{r}), 	\label{Atr}
\end{eqnarray}
Hamiltonian $\hat{H}_0$ after the substitution (\ref{subst}) also satisfies theorem 1.

The only question remained is whether any admissible distribution $\mathbf{h}(\mathbf{r})$ has the corresponding vector potential with the proper translational symmetry. The answer is not obvious because, for instance, the vector potential corresponding to a homogeneous magnetic field is not translationally invariant at any gauge. 

To define the vector potential
for given $\mathbf{h}(\mathbf{r})$ one can apply the Coulomb gauge, which leads to the following equation \cite{Knoepfel}   
\begin{eqnarray}
	\Delta \mathbf{A} = - \mu_0 \mathbf{j}
	\label{vectpot}
\end{eqnarray}
where $\mu_0$ is the magnetic constant and $\mathbf{j} =\nabla \times \mathbf{h}$ is the current density, which generates the magnetic field $\mathbf{h}$. In 3D Cartesian coordinates
this equation splits into three independent Poisson equations \cite{Knoepfel}: $	\Delta A_i = - \mu_0 j_i$, where $i=x,y,z$. Since $\mathbf{j}(\mathbf{r}+\mathbf{T/2})= -\mathbf{j}(\mathbf{r})$, the solution of Eq.~(\ref{vectpot}) satisfies the relation (\ref{Atr}) for periodic or homogeneous boundary conditions. The above-mentioned example of homogeneous magnetic field is a solution with inhomogeneous boundary conditions at infinitely distant boundaries (surface currents at the boundaries). 

In solids, one can consider $\mathbf{j}$ as a molecular current which produces the helical magnetic field. That is why, theorem 1 can be extended to the charge interaction with magnetic field (\ref{subst}) for any reasonable distribution $\mathbf{h}(\mathbf{r})$.

Helical systems with the SSG symmetry operator $\{\mathbf{ \boldsymbol\alpha}|0|\mathbf{t}\}$, where $\alpha=2\pi/n$ and $n$ is an even natural number, satisfy the hypothesis of the theorem 1 if $\mathbf{h}(\mathbf{r})$ is perpendicular to the axis of spin rotation as illustrated by a tight-binding model for a four-sublattice helical structure in the next subsection. On the other hand, the theorem can not be applied if $n$ is the odd number, in particular, in the important case of the 120$^0$ magnetic ordering ($n=3$).

Let us introduce an operator $\hat{\mathbf{r}}_{\alpha}$ of spin rotation by angle $\alpha$ about $z$ axis. In the theorem and models discussed below, there is no spin-orbit coupling. Therefore, the spin rotation axis can be chosen arbitrary, i.e. it is independent of the spacial $z^\prime$ axis.

{\bf{Theorem 2.}} If $\hat{H}_0$ in Eq.(\ref{H}) is invariant under translation $\hat{\mathbf{t}}$, time reversion $\hat{\mathbf{\theta}}$, and arbitrary rotations of the spin system about $z$ axis, and if $\mathbf{h}(\mathbf{r}) \hat{\mathbf{\boldsymbol\sigma}}$ is invariant under $\hat{\mathbf{r}}_{\alpha} \hat{\mathbf{t}}$, where  $\alpha=2\pi/n$, $n$ is an odd number ($n>1$), and $\mathbf{h}(\mathbf{r})$ is perpendicular to the spin rotation axis, then the eigenvalues of $\hat{H}$ are at least two-fold degenerate for all the wave vectors $\mathbf{k}$ except  those, which satisfy
\begin{eqnarray}
	\exp\big(2 i \mathbf{k} \mathbf{T}\big)=1, \label{exp2}
\end{eqnarray}
where $\mathbf{T}=n\mathbf{t}$, and
all the eigenvalues satisfy Eq.~(\ref{T1}) with $\langle \hat{\mathbf{\sigma}}_{x(y)} \rangle_ {\mathbf{k}} = 0$.

{\bf{Proof.}}  Let us introduce operators    
\begin{eqnarray}
	\hat{\mathbf{Y}} = \hat{\mathbf{t}} \hat{\mathbf{r}}_{\alpha}  \mbox{   and   }	\hat{\mathbf{X}} = \hat{\mathbf{t}} \hat{\mathbf{r}}_{\alpha-\pi} \hat{\mathbf{\theta}}. \label{XY}
\end{eqnarray}
$\hat{\mathbf{Y}}$ is a symmetry operator under the hypothesis of the theorem and a generator of an Abelian group \cite{Sandratskii}. Its irreducible representations coincide with those of the space translation group $\exp(i\mathbf{k}\mathbf{t})$,
where $\mathbf{k}$ is the wave vector in the extended Brillouin zone \cite{Sandratskii}. The magnetic field is perpendicular to the spin rotation axis, then $\hat{\mathbf{r}}_{-\pi}\mathbf{h}(\mathbf{r}) = - \mathbf{h}(\mathbf{r})$ and $\hat{\mathbf{X}}$ is also a symmetry operator. We can consider the following quantity:
\begin{eqnarray}
	\langle \psi_ {\mathbf{k}} | \big( \hat{\mathbf{X}}^n | \psi_ {\mathbf{k}} \rangle \big) = 	\langle \psi_ {\mathbf{k}} | \big( \hat{\mathbf{X}}^{-n} | \psi_ {\mathbf{k}} \rangle \big).  \label{H21}
\end{eqnarray}
Forward transformations give
\begin{eqnarray}
	\langle \psi_ {\mathbf{k}} | \big( \hat{\mathbf{t}}^{-n} \hat{\mathbf{r}}_{\alpha-\pi}^{-n} \hat{\mathbf{\theta}}^{-n} | \psi_ {\mathbf{k}} \rangle \big) =  \langle \psi_ {\mathbf{k}} | \hat{\mathbf{t}}^{-2n} \big( \hat{\mathbf{t}}^{n} \hat{\mathbf{r}}_{\alpha-\pi}^{-n} \hat{\mathbf{\theta}}^{-n} | \psi_ {\mathbf{k}} \rangle \big) \nonumber \\
	=  - \langle \psi_ {\mathbf{k}} | \hat{\mathbf{t}}^{-2n} \hat{\mathbf{r}}_{\alpha-\pi}^{-2n} \big( \hat{\mathbf{t}}^{n} \hat{\mathbf{r}}_{\alpha-\pi}^{n} \hat{\mathbf{\theta}}^{n} | \psi_ {\mathbf{k}} \rangle \big) .  \label{H22}
\end{eqnarray}
Using $\hat{\mathbf{r}}_{\alpha-\pi}^{-2n} = - \hat{\mathrm{I}}$ and Bloch's theorem $ \hat{\mathbf{T}}^2| \psi_ {\mathbf{k}} \rangle = \exp\big(2 i \mathbf{k} \mathbf{T}\big)| \psi_ {\mathbf{k}} \rangle$, we obtain
\begin{eqnarray}
	\langle \psi_ {\mathbf{k}} | \big( \hat{\mathbf{X}}^n | \psi_ {\mathbf{k}} \rangle \big) = \exp\big( - 2 i \mathbf{k} \mathbf{T}\big)	\langle \psi_ {\mathbf{k}} | \big( \hat{\mathbf{X}}^n | \psi_ {\mathbf{k}} \rangle \big).  \label{H23}
\end{eqnarray}
That is, $| \psi_ {\mathbf{k}} \rangle$ and $\hat{\mathbf{X}}^n | \psi_ {\mathbf{k}} \rangle$ are orthogonal and the eigenstates are two-fold degenerate for all the wave vectors except those, which satisfy Eq.~(\ref{exp2}).

The operators $\hat{\mathbf{k}}$ and $\hat{\sigma}_z$ commute with $\hat{\mathbf{t}}$ and $\hat{\mathbf{r}}$, as well as they change sign under the time reversal. Therefore, we obtain 
$\varepsilon_{\mathbf{k},\langle \sigma _z \rangle} = \varepsilon_{-\mathbf{k}, - \langle \sigma _z \rangle}$.

The relations for transverse spin components can be proven by direct calculation. The average of spin projection on $x$ axis is defined by
\begin{eqnarray}
	\langle \hat{\mathbf{\sigma}}_{x} \rangle_ {\mathbf{k}} = \frac{1}{V} \int_V \varphi^*_ {\mathbf{k}}(\mathbf{r}) \hat{\mathbf{\sigma}}_{x} \varphi_ {\mathbf{k}}(\mathbf{r}) d\mathbf{r} \label{Sx}
\end{eqnarray}
where the integration is performed over the magnetic unit cell and $V$ is its volume. However, there are alternative choices of the magnetic unit cell, e.g. one can shift it by means of generalized translations, i.e. by $\hat{\mathbf{t}} \hat{\mathbf{r}}_{\alpha}$ or $\hat{\mathbf{-t}} \hat{\mathbf{r}}_{-\alpha}$. The only effect of these operations is the spin rotation about $z$ axis:
\begin{eqnarray}
	\langle \hat{\mathbf{\sigma}}_{x} \rangle_ {\mathbf{k}}^{\prime} =\cos(\alpha) \langle \hat{\mathbf{\sigma}}_{x} \rangle_ {\mathbf{k}} + \sin(\alpha) \langle \hat{\mathbf{\sigma}}_{y} \rangle_ {\mathbf{k}}
	 \nonumber \\
	\langle \hat{\mathbf{\sigma}}_{x} \rangle_ {\mathbf{k}}^{\prime \prime} =\cos(\alpha) \langle \hat{\mathbf{\sigma}}_{x} \rangle_ {\mathbf{k}} - \sin(\alpha) \langle \hat{\mathbf{\sigma}}_{y} \rangle_ {\mathbf{k}}. \label{Sxy}
\end{eqnarray}
Since $\hat{\mathbf{t}} \hat{\mathbf{r}}_{\alpha}$ is a symmetry operation the average of spin projection should be independent on the particular choice of the cell, i.e. $\langle \hat{\mathbf{\sigma}}_{x} \rangle_ {\mathbf{k}} = \langle \hat{\mathbf{\sigma}}_{x} \rangle_ {\mathbf{k}}^{\prime} = \langle \hat{\mathbf{\sigma}}_{x} \rangle_ {\mathbf{k}}^{\prime \prime}$. It is easy to see that the only solution of the system of equations is $\langle \hat{\mathbf{\sigma}}_{x} \rangle_ {\mathbf{k}} = 0$ (apart from the trivial solution $\cos(\alpha)=1$). The identity $\langle \hat{\mathbf{\sigma}}_{y} \rangle_ {\mathbf{k}} = 0$ can be proved similarly. $\blacksquare$.

\subsection{1D examples of the theorems applications}

To illustrate the theorems proved above let us consider a tight-binding model of a simple atomic chain under an effective magnetic field:
\begin{eqnarray}
	\hat{H}_{1D}=	-\sum_{i,\sigma}\left[ \sum_{j=1}^{j_m -1}\left( \hat{a}^\dagger_{i,j,\sigma} \hat{a}_{i,j+1,\sigma} +h.c. \right) + \right. \nonumber\\ \left. \left( \hat{a}^\dagger_{i,j_m,\sigma} \hat{a}_{i+1,1,\sigma} + h.c. \right)  -
	\sum_{j,\sigma^\prime}\left( \hat{a}^\dagger_{i,j,\sigma} \hat{\mathbf{h}}_{j} \hat{a}_{i,j,\sigma^\prime} \right) \right].  \label{H3}
\end{eqnarray}
where $\hat{a}^\dagger_{i,j,\sigma} (\hat{a}_{i,j,\sigma} )$ is the electron creation (annihilation) operator in the $j$-th sublattice ($j=1, \ldots ,j_{m}$) and $i$-th cell with the spin projection on $z$ axis $s_z=\pm 1/2$, $\hat{\mathbf{h}}_{j} = \mathbf{h}_j \hat{\mathbf{ \boldsymbol\sigma}}$.

\begin{figure}	\includegraphics[width=0.45\textwidth]{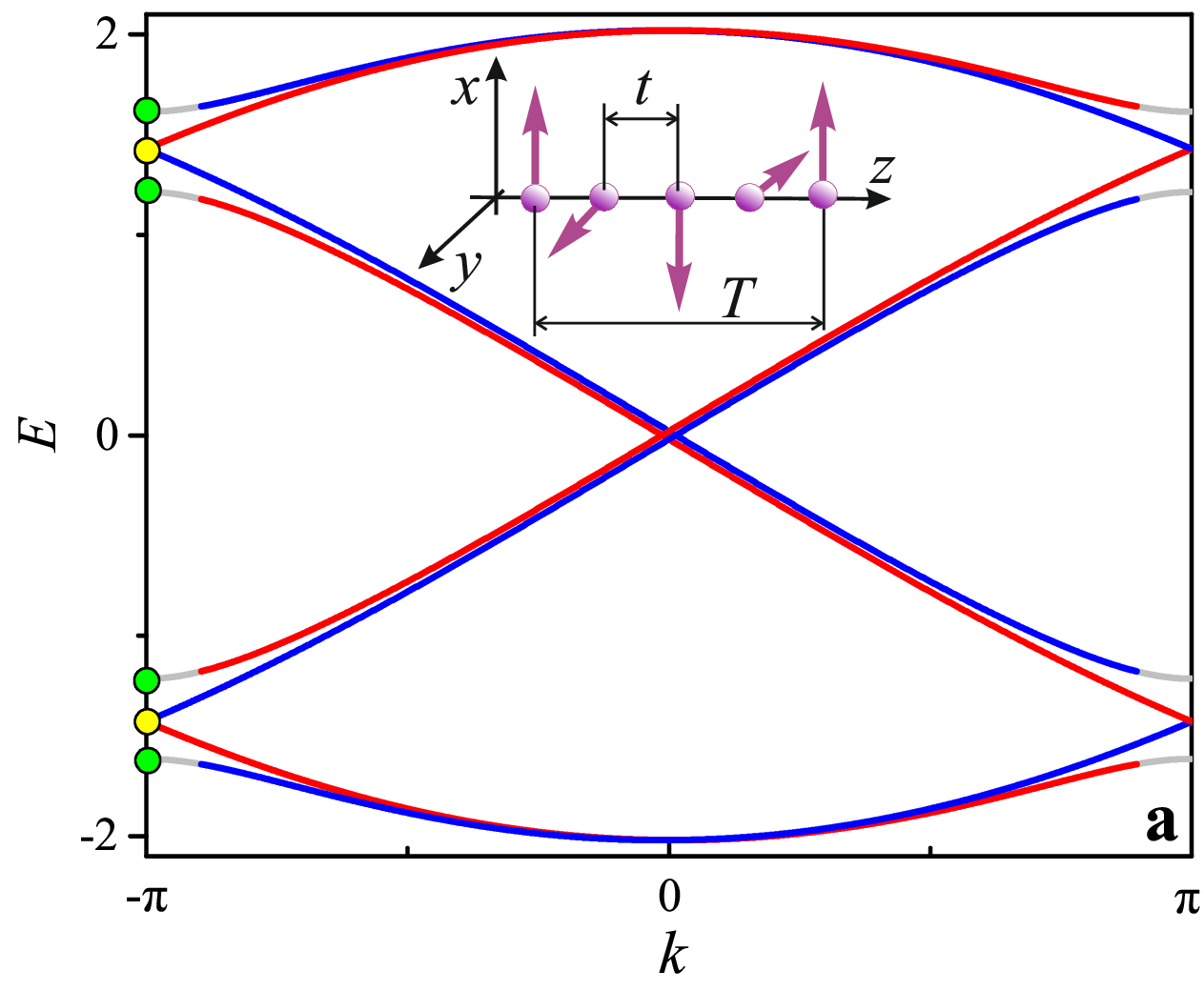} \\
	\includegraphics[width=0.45\textwidth]{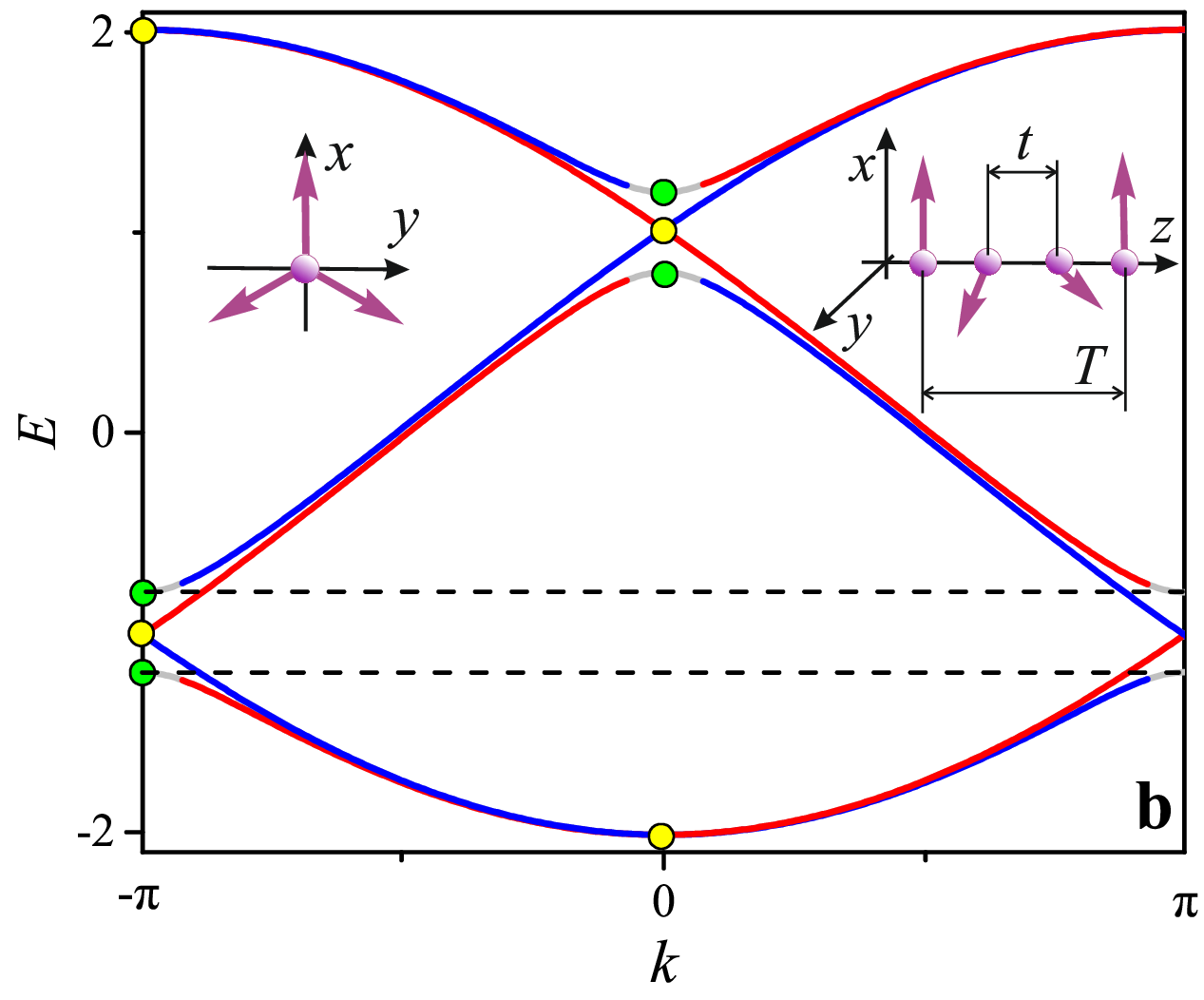} 
	\caption{\label{f1} (a) The electron band structure of the 1D model Eq.~(\ref{H3}) and (b) a schematic cylindrical representation of the bottom dispersion curve, which is shown by the thick line in (a) panel.  The average spin along the curves is indicated by color: red and blue if $|\langle \hat{\mathbf{\sigma}}_z\rangle|>1/2$, gray otherwise. The yellow and green circles denote the degenerate and non-degenerate points, correspondingly.}
\end{figure}

We start with a four sublattice version of the model ($j_m=4$) which satisfies theorem 1: $\hat{\mathbf{h}}_{1}=h_0 \hat{\sigma}_x$, $\hat{\mathbf{h}}_{2}=h_0 \hat{\sigma}_y$, $\hat{\mathbf{h}}_{3}= - h_0 \hat{\sigma}_x$, and $\hat{\mathbf{h}}_{4}= - h_0 \hat{\sigma}_y$. This model is exactly solvable \cite{KudasovPRB}. The electron dispersion and schematic representation of this system are shown in Fig.~\ref{f1}a. The magnetic Brillouin zone lies between $-\pi$ and $\pi$. 
According to theorem 1, the obtained eigenvalues obey Eq.~(\ref{T1}) where spin states are shown by color. In the special points, where the degeneracy is undefined ($k=\pm\pi$), degenerated and nodegenerated states are shown by yellow and green circles.
It is worth noticing that the dispersion in Fig.~\ref{f1}a corresponds to the $p$-wave antiferromagnet \cite{Brekke}.

A three sublattice model ($j_m=3$) corresponding to the 120$^0$ order meets the condition of theorem 2. The on-site magnetic field for the sublattices is defined as follows: $\hat{\mathbf{h}}_{1}=h_0 \hat{\sigma}_x$, $\hat{\mathbf{h}}_{2}=h_0 (-\hat{\sigma}_x+\sqrt{3}\hat{\sigma}_y)/2$, $\hat{\mathbf{h}}_{3}=h_0 (-\hat{\sigma}_x-\sqrt{3}\hat{\sigma}_y)/2$. This model is also exactly solvable. The schematic view of magnetic structure and electron dispersion are shown in Fig.~\ref{f1}b. The symmetry condition  Eq.~(\ref{T1}) is also  fulfilled. In contrast to the four sublattice model there are two types of special points, where the degeneracy is undefined ($k=0$ and $k=\pm\pi$). They are also shown in Fig.~\ref{f1}b  by color circles.

\subsection{2D model with 120$^0$ magnetic ordering}

\begin{figure}
	\includegraphics[width=0.45\textwidth]{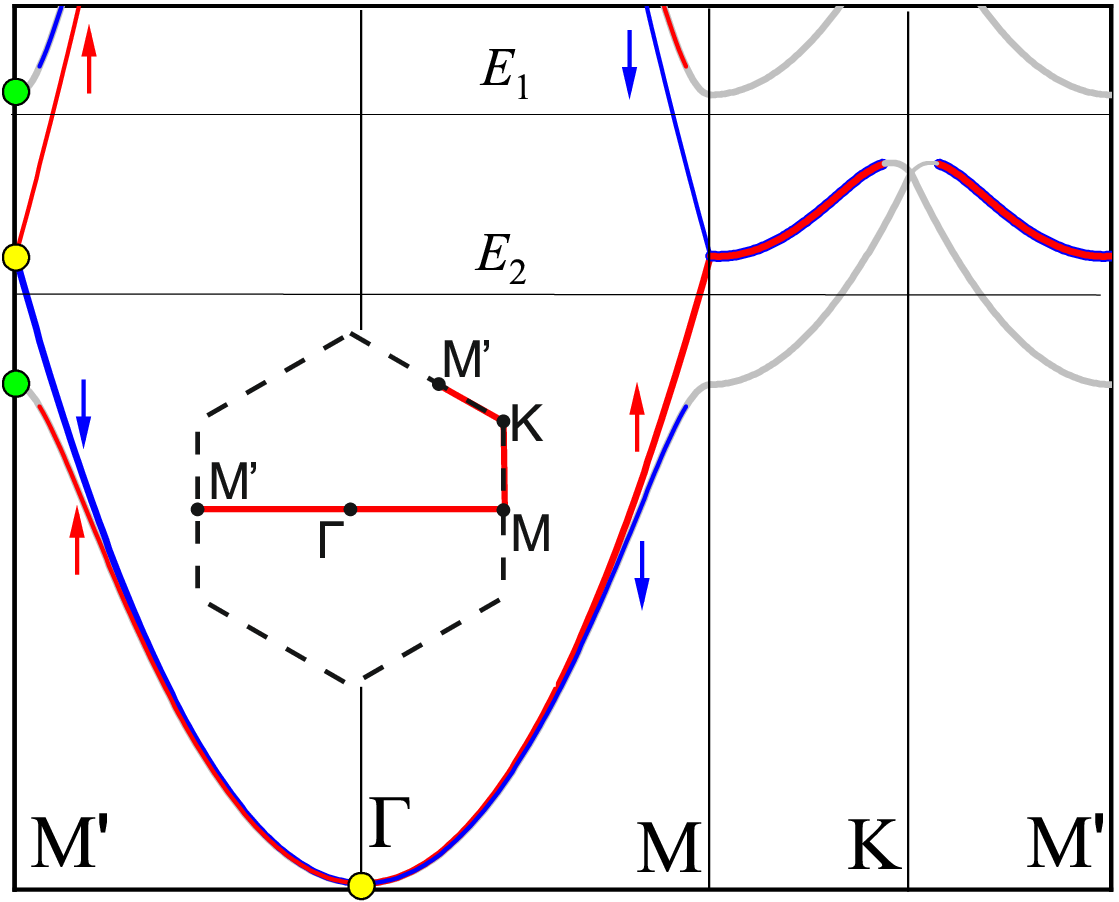}
	\caption{\label{f2} The dispersion of nearly-free electron model Eq.~(\ref{NFE}) along the path shown in the insert. The colors of lines and circles are defined in the same manner as in Fig.~\ref{f1}.}
\end{figure}

The proved theorems are also valid for multidimensional systems. Below we discuss three-sublattice model on a 2D lattice, which will be used for an investigation of conductive hexagonal palladium layers in PdCrO$_2$. They are described well by a 2D nearly-free-electron model \cite{Kushwaha}. Magnetic spacers CrO$_2$ form an effective field corresponding to a 120$^0$ (three-sublattice) magnetic structure \cite{Takatsu2}. Then, a solution for the Bloch wave functions
\begin{eqnarray}
\hat{\psi}_{\mathbf{k}}= \sum_{\mathbf{K}} \hat{c}_{\mathbf{k} - \mathbf{K}} \exp\left[ i (\mathbf{k}-\mathbf{K}) \mathbf{r}  \right]
\label{psi}
\end{eqnarray}
can be obtained from the following equation \cite{Ashcroft}
\begin{eqnarray}
	\left[ \frac{\hbar^2}{2m} \big(\mathbf{k} - \mathbf{K}\big)^2 - \mathcal{E} \right] \hat{c}_{\mathbf{k} - \mathbf{K}} +\sum_{\mathbf{K^\prime}} \hat{U}_{\mathbf{\mathbf{K^\prime}} - \mathbf{K}} \hat{c}_{\mathbf{k} - \mathbf{K^\prime}} = 0 \; \; \; \; \label{NFE}
\end{eqnarray} 
where $\mathbf{k}$ is the wave vector within the magnetic Brillouin zone ($\sqrt3 \times \sqrt3$), $\hat{U}_{\mathbf{K}}$ are the Fourier coefficients of effective field, $\hat{c}_{\mathbf{k}}$ and $\hat{U}_{\mathbf{K}}$ have spinor form \cite{Kudasov}. An example of a continuous spinor field corresponding to the three-sublattice structure was discussed previously \cite{KudasovFTT}. A general form of $\hat{U}_{\mathbf{K}}$ for helical spinor field is derived in Appendix A. In case of 2D system, the second term in Eq.~(\ref{NFE}) should contain at least two terms to describe properly the band structure in the vicinity of the K points (see Fig.~\ref{f2}).

The band structure of 2D model is shown in Fig.~\ref{f2}. It is easy to see that the Eq.~(\ref{T1}) is valid. In the 1D model the condition defined by Eq.~(\ref{exp2}) led to isolated points (Fig.~\ref{f1}b). In case of 2D model it gives isolated lines, which coincide with boundaries of the magnetic Brillouin zone, and an isolated point (the $\Gamma$ point in Fig.~\ref{f2}). They are shown in Fig.~\ref{f2} by the colored circles.

\section {Topology of dispersion surfaces}

\subsection{Generalized Bloch theorem and periodic boundary conditions}

Let us consider an electron motion in a periodic potential of crystal lattice $U (\mathbf{r} + \mathbf{t}) = U (\mathbf{r})$, where $\mathbf{t}$ is any of the primitive translation vector, and a commensurate helical magnetic field. Thus $\hat{H}_0 = -\nabla^2 + U (\mathbf{r})$ in Eq.~(\ref{H}), and  $\hat{ \mathbf{r}}_{i} \hat{\mathbf{t}}_{i} \: \mathbf{h}(\mathbf{r})  = \mathbf{h}(\mathbf{r})$, where $\hat{\mathbf{t}}_{i}$ is the $i$-th  primitive vector translation operator,  $\hat{\mathbf{ \mathbf{r}}}_{i}$ is the rotation of the spin system about $z$ axis by the angle $\alpha_i = 2 \pi / N_i$. If $N_i$ is an even number the present model satisfy theorem 1. In case of $N_i$ being odd, the model includes systems, which meet the conditions of theorem 2, but goes far beyond it because the magnetic field is not necessarily perpendicular to the  $z$ axis.   

Under these assumptions Hamiltonian (\ref{H}) is invariant with respect to the generalized translation $ \hat{\mathbf{ \mathbf{r}}}_{i} \hat{\mathbf{\mathbf{t}}}_{i} $, which corresponds to the SSG operator $\{\mathbf{ \boldsymbol\alpha}_i| 0 |\mathbf{t}_i\}$. It forms a cyclic group, which is isomorphic to that of ordinary translations \cite{Sandratskii}. Its irreducible representation has the following form $D_{\mathbf{k}}(\hat{\mathbf{ \mathbf{r}}}_{i} \hat{\mathbf{\mathbf{t}}}_{i}) = \exp(i \mathbf{k} \mathbf{t}_{i})$. Then an eigenfunction of Hamiltonian (\ref{H}) can be written down as
\begin{eqnarray}
	\psi_ {\mathbf{k}\sigma}(\mathbf{r}) = \exp(i \mathbf{k} \mathbf{r}) 	\hat{u}_\mathbf{k}(\mathbf{r}), \label{Bloch}
\end{eqnarray}
where the spinor function $\hat{u}_\mathbf{k}(\mathbf{r})$ possesses a generalized periodicity, i.e.
\begin{eqnarray}
	\hat{\mathbf{ \mathbf{r}}}_{i} \hat{\mathbf{\mathbf{t}}}_{i} \hat{u}_\mathbf{k}(\mathbf{r}) = \hat{u}_\mathbf{k}(\mathbf{r}). \label{rt}
\end{eqnarray}
 Here, the wave vector $\mathbf{k}$ is defined in crystallographic (extended) Brillouin zone. This statement is the generalized Bloch theorem for helical magnetic systems \cite{Sandratskii}.

After the substitution of Eq.(\ref{Bloch}) to the Hamiltonian one can obtain the equation for  $\hat{u}_\mathbf{k}(\mathbf{r})$:
\begin{eqnarray}
	[ (-i\nabla + \mathbf{k})^2 + U( \mathbf{r})+\mathbf{h}(\mathbf{r}) \hat{\mathbf{ \boldsymbol\sigma}}] \hat{u}_\mathbf{k}(\mathbf{r}) =\mathcal{E}_\mathbf{k} \hat{u}_\mathbf{k}(\mathbf{r}). \label{Hk}
\end{eqnarray}
From here, one can see that the dispersion $\mathcal{E}_\mathbf{k}$ should be continuous and periodic within the {\it{crystallographic Brillouin zone}}.  

To define the band structure we should establish the periodic boundary condition of a Born–-von Karman type in addition to the Bloch theorem. They have the following form
\begin{eqnarray}
	\psi_ {\mathbf{k}\sigma}(\mathbf{r}+N_i \mathbf{T}_i) = \psi_ {\mathbf{k}\sigma}(\mathbf{r}), \label{BornKarm}
\end{eqnarray}
where $ \mathbf{T}_i $ are the primitive vectors of magnetic lattice, which is defined by the crystallographic translations: $ \mathbf{T}_i = \sum_{ij} n_{ij} \mathbf{t}_j$ where $n_{ij}$ are integer numbers. A magnetic unit cell has the volume $ V_m $ which is a multiple of the volume of crystallographic unit cell $V_c$, i.e. $ N_m = V_m/V_c $.

It follows from the boundary conditions that the allowed values of wave vectors lie within the {\it{magnetic Brillouin zone}}. That is, the dispersion  $\mathcal{E}_\mathbf{k}$ defined by Eq.~(\ref{Hk}) should be reduced to it.
Then, a band structure with $N_m$ branches appears. It has the following properties. (i) The structure is gapless since the initial dispersion $\mathcal{E}_\mathbf{k}$ is continuous. (ii) An individual branch is not necessarily periodic in the magnetic Brillouin zone while the entire band structure remains periodic according to the ordinary Bloch theorem. This leads to profound consequences in topology of the band structure as shown in the examples in the next subsections.

\subsection{Topology of 1D dispersion curves}

\begin{figure*}	\hfill \includegraphics[width=0.55\textwidth]{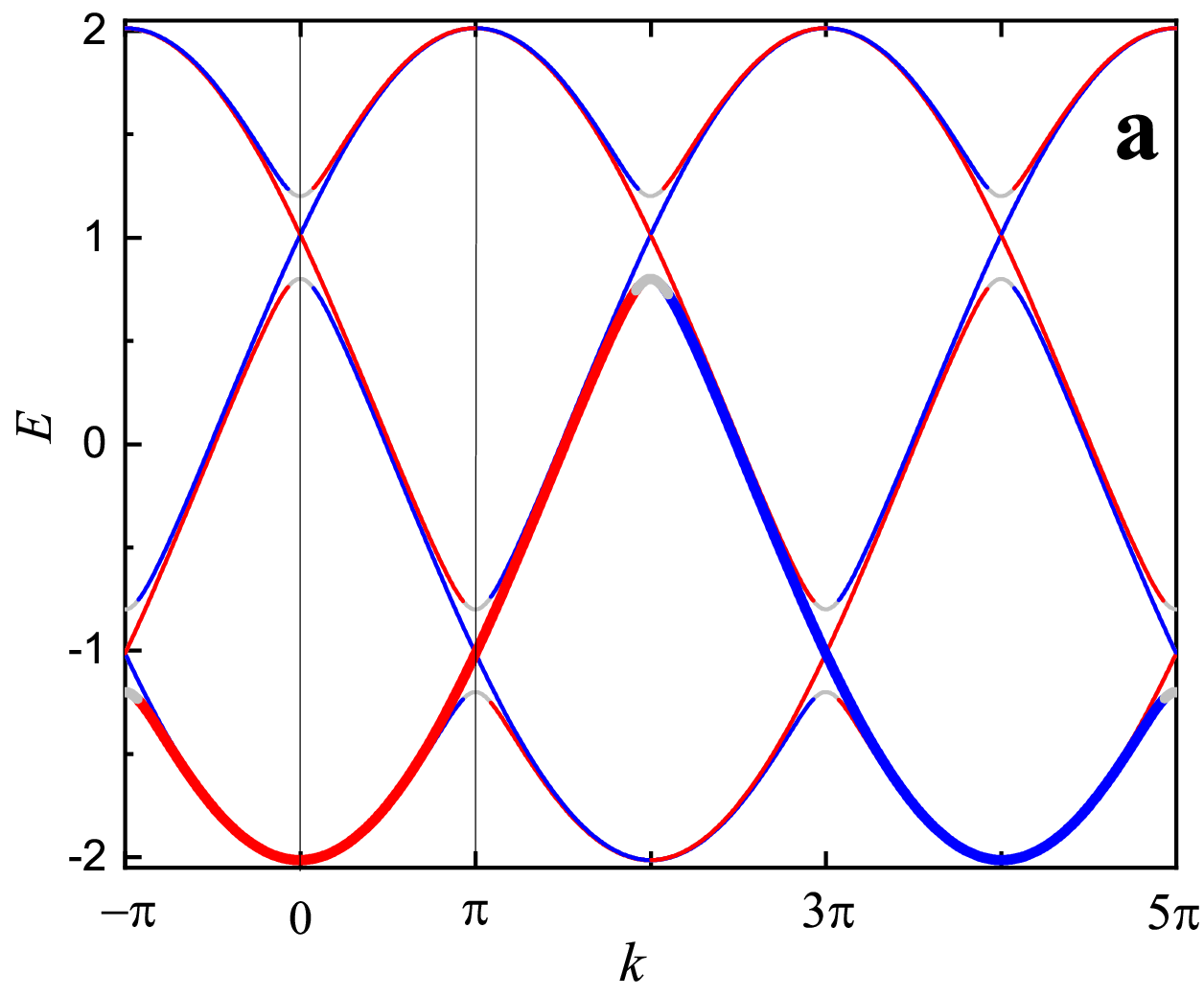} \hfill
	\includegraphics[width=0.3\textwidth]{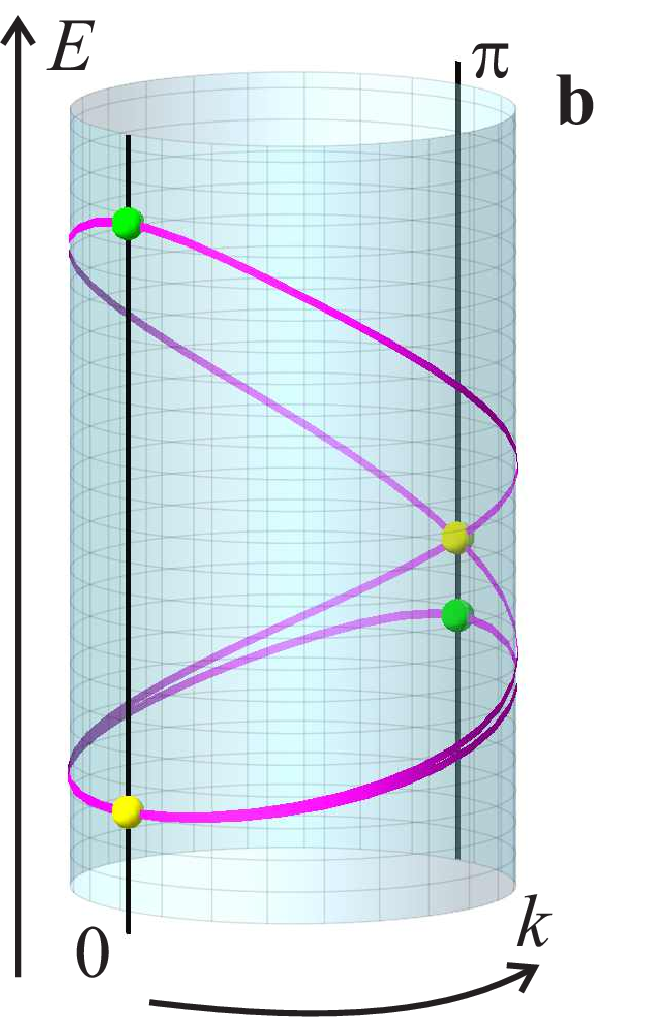}\hfill
	\caption{\label{f3} (a) The electron band structure of the 1D model Eq.~(\ref{H3}) and (b) a schematic cylindrical representation of the bottom dispersion curve, which is shown by the thick line in (a) panel \cite{KudasovPRB}.  The average spin along the curves in panel (a) is indicated by color in the same manner as in Fig.~\ref{f1}. The yellow and green spheres denote the degenerate and non-degenerate points, correspondingly.}
\end{figure*}

The band structure of the three sublattice 1D model discussed above as an example for application of theorem 2 is shown in Fig.~\ref{f1}b. Let us extend it to the crystallographic Brillouin zone (Fig.~\ref{f3}a). One can see that the three lower branches in Fig.~\ref{f3}a in fact form a continuous periodic dispersion in the extended Brillouin zone and non-periodic in the magnetic one according to the discussion in previous subsection ($N_m =3$).    

To investigate the topology of 1D band structure we represent the dispersion curves as a line on cylinder, i.e. in the space $S^1\times E^1$ where $S^1$ and $E^1$ are the unit circle and unit interval, correspondingly. An ordinary dispersion curve, which is periodic in the magnetic Brillouin zone, is homeomorphic to a circle $S^1$.
The curve shown in Fig.~\ref{f3}b has $N_m$ turns. There are also self-crossings. At the same time, if we threat the curve as a line in a full state space, for instance $S^1\times S^{n}$ and $n>1$, the self-crossing disappear because eigenvectors of the Hamiltonian at a fixed value of wave vector are orthogonal, and, therefore, they cannot have the same sets of independent parameters.  

The fundamental group of the state space $\pi_1 (S^1\times S^{n})$ can be easily calculated since the fundamental group of a direct product of two spaces is isomorphic to a product of its fundamental groups and $\pi_1 (S^{n})$ at $n>1$ is the unit group. That is, $(S^1\times S^{n}) \sim \pi_1 (S^1)$, and we obtained the fundamental group of circle. This means that all closed lines (paths) on cylinder can be divided into classes depending on winding number $N_m$ and direction of the path \cite{Malcev}. These classes are homotopically nonequivalent, that is, paths from different classes cannot be transformed one into another by smooth deformations. Therefore, the dispersion curve shown in Fig.~ by the thick line is topologically nontrivial, and $N_m$ is a topological invariant.

The topology of dispersion curve is stable against variations of the periodic potential $U (\mathbf{r})$. To investigate an effect of homogeneous magnetic field it is useful to define a unit chirality vector $\mathbf{n}_{\chi}$ (see Appendix B) and projection of the magnetic field on to it, $B_{\chi} = \mathbf{n}_{\chi} \mathbf{B}$. Then, an additional homogeneous magnetic field with the only component $B_{\chi}$ also does not affect the topology because the generalized periodicity of the Hamiltonian is conserved. At the same time, the conditions of theorems are violated and, therefore, Eq.~(\ref{T1}) is no longer valid. If the homogeneous magnetic field lies in the spin plane ($\mathbf{n}_{\chi} \perp \mathbf{B}$) it breaks the rotational symmetry of the helicoid, and the topology of band structure becomes trivial \cite{Kudasov}.    

\subsection{Topology of 2D dispersion surfaces}

Let us investigate the dispersion curves of the 2D model in detail. According to the generalized Bloch theorem and periodic boundary conditions, the allowed wave vectors lie in the magnetic Brillouin zone, however the dispersion is continues and periodic in the crystallographic one, that is, in $\sqrt{3} \times \sqrt{3}$ cell rotated by $\pi /6$. As a result the curves in Fig.~\ref{f2} are non-periodic, that is, they transit one to another while crossing boundaries of the magnetic Brillouin zone, and at the same time they are continuous and periodic in crystallographic Brillouin zone. 

It is convenient to investigate the topology of dispersion surfaces within a rhombic representation of the magnetic Brillouin zone. To take into account the periodical boundary conditions, we can glue its opposite edges and map it onto
the surface of torus $S^1 \times S^1$  \cite{Tkachev}. The state space has the following form:  $S^1 \times S^1 \times S^{n}$ ($n>1$) \cite{KudasovJETPL2}.

\begin{figure}
	\includegraphics[width=0.2\textwidth]{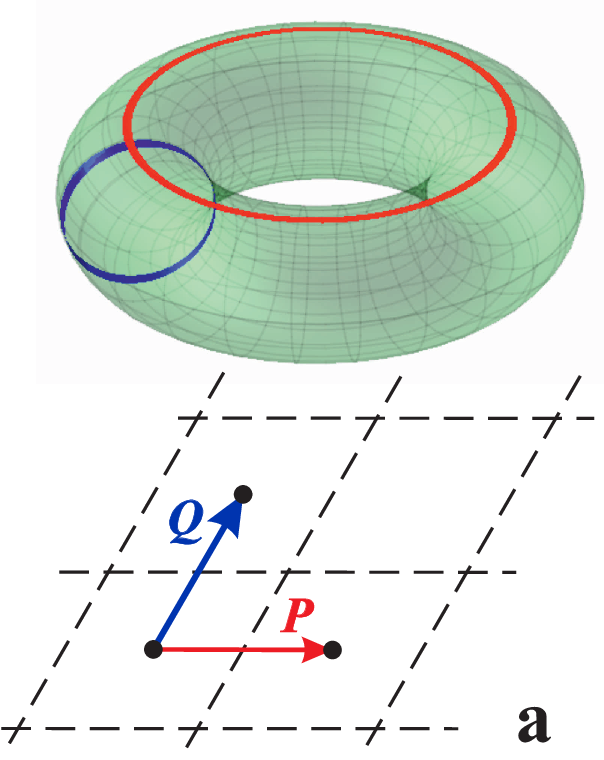} \hfill
	\includegraphics[width=0.26\textwidth]{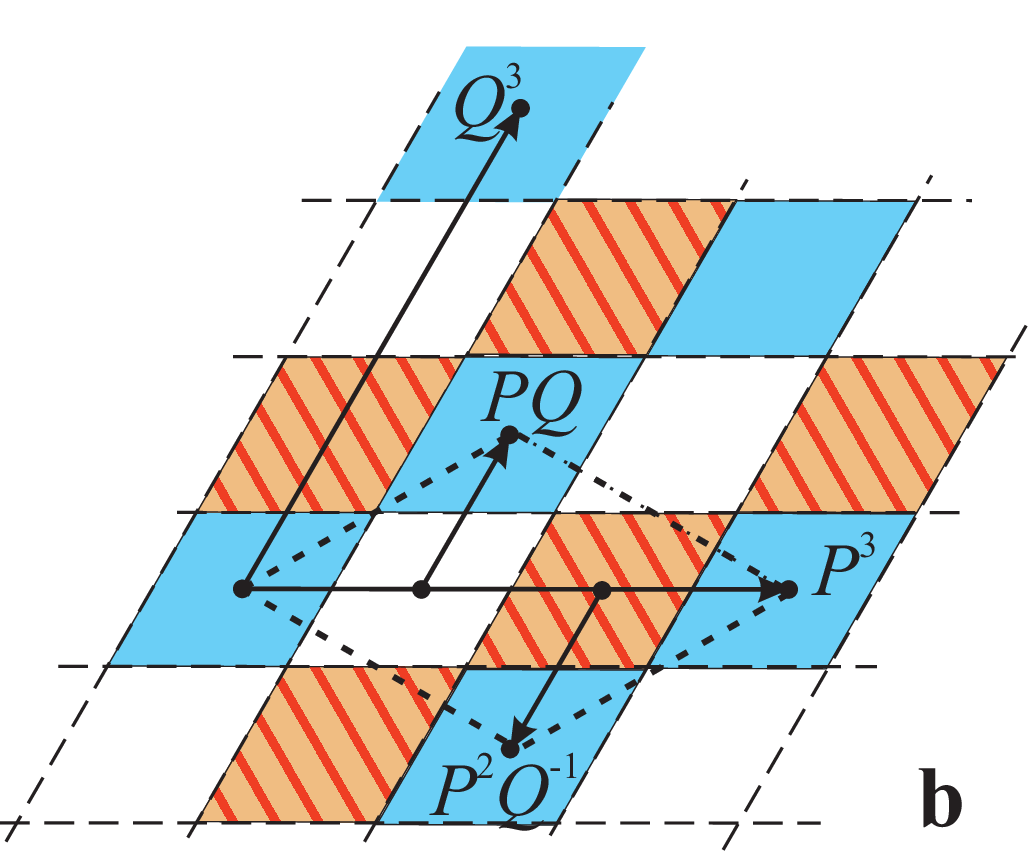}
	\caption{\label{f4} A schematic representation of electron dispersion in the rhombic magnetic Brillouin zone for (a) single-sheet (together with a torus view) and (b) multi-sheet surfaces \cite{KudasovJETPL2}.}
\end{figure}

To elucidate the topology of dispersion surface let us investigate its fundamental group \cite{Massey,Dubrovin}. An ordinary dispersion surface is homeomorphic to the toroidal one $S^1 \times S^1$. The planar and toroidal representations of the magnetic Brillouin zone are shown in Fig.~\ref{f4}a. There are two closed nonzero primitive paths, which cannot be contracted to a point: $P$ and $Q$. They encircle the torus in one of two ways. In the planar representation they intersect horizontal or tilt boundary of the magnetic Brillouin zone. Since any closed path without a boundary crossing can be contracted to a point, an exact positions of the initial and final points are inessential, and it is assumed hereinafter that they are at the zone center ($\Gamma$ point). Then, an arbitrary path on ordinary dispersion surface has the form $P^n Q^m$ where $n$ and $m$ are integers. It is a representation of the torus fundamental group \cite{Malcev}.

A multisheet dispersion surface is a covering space with respect to an ordinary one. The three sheets of the surface discussed above is shown in Fig.~\ref{f4}b by colors and texture. It is easy to see that $P$ and $Q$ are open paths because the initial and final points lie in the different sheets. Minimal closed paths are $P Q$, $P^2 Q^{-1}$, $P^3$, and $Q^3$. Nevertheless, after the substitution $A = P Q$ and $B = P^2 Q^{-1}$ we can represent an arbitrary closed path as $A^n B^m$, i.e. the fundamental group of the multisheet dispersion surface is isomorphic to that of torus. It should be mentioned that if the paths $A$ and $B$ are used to build a unit cell in the $\mathbf{k}$-space as shown in Fig.~\ref{f4}b by the dash lines, then we obtain the crystallographic Brillouin zone. 

The fundamental group of the state space is defied similarly to that of the 1D model and can be reduced to to the fundamental group of torus $\pi_1 (S^1\times S^1\times S^{n}) \sim \pi_1 (S^1\times S^1)$. As can be seen from above, the both types of dispersion surfaces are embeddings of a torus in the state space $S^1 \times S^1 \times S^{n}$. However, closed paths $P$ and $Q$ exist on an ordinary surface and do not on a multisheet one. That is, these surfaces are not isotopic \cite{note} to each other, i.e., there is no
continuous deformation of the space (homotopy) connecting them to each other \cite{Dubrovin,Kalajdzievski}. That is why, a multisheet dispersion surfaces is not topologically equivalent to an ordinary one, and the number of covering sheets ($N_m=3$), which is equal to the ration of volumes of crystallographic and magnetic Brillouin zones, is a topological invariant. It should be stressed that the isotopic equivalence between the two types of the dispersion surfaces is broken by the topology of the underlying state space $S^1 \times S^1 \times S^{n}$.

Similarly to the 1D model, a homogeneous magnetic field, which is along the chirality vector ($\mathbf{n}_{\chi} \parallel \mathbf{B}$), conserves the band structure topology, and that lying in the plane ($\mathbf{n}_{\chi} \perp \mathbf{B}$) brings the band structure to the trivial topological state.          

\section{Transport properties in topological metals}

\subsection{Fermi surface for 2D model}

In the previous sections it was shown that the dispersion curves (surfaces) in metallic helimagnets have the specific symmetry (Eq.~(\ref{T1})) and can be non-periodic within the magnetic Brillouin zone. This allows for the existence of a single band crossing the Fermi level, which is asymmetric in spin.
For instance, it occurs if the Fermi level fall into the range between dash lines in Fig.~\ref{f1}b. The electron transport becomes unusual in this case \cite{Kudasov}: there is no backward scattering without spin-flip, and a permanent spin current can arise.

\begin{figure}
 	\includegraphics[width=0.21\textwidth]{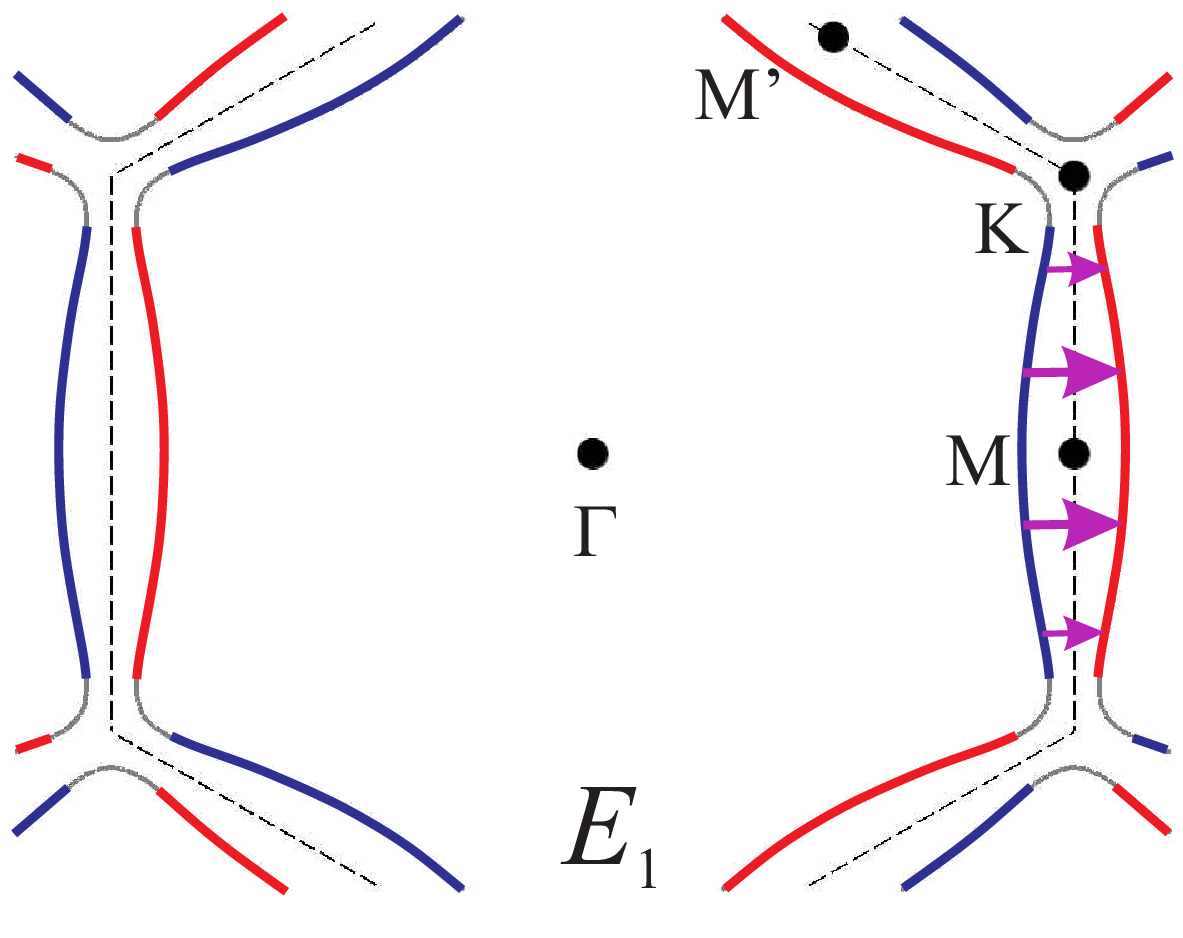} \hfill
	\includegraphics[width=0.21\textwidth]{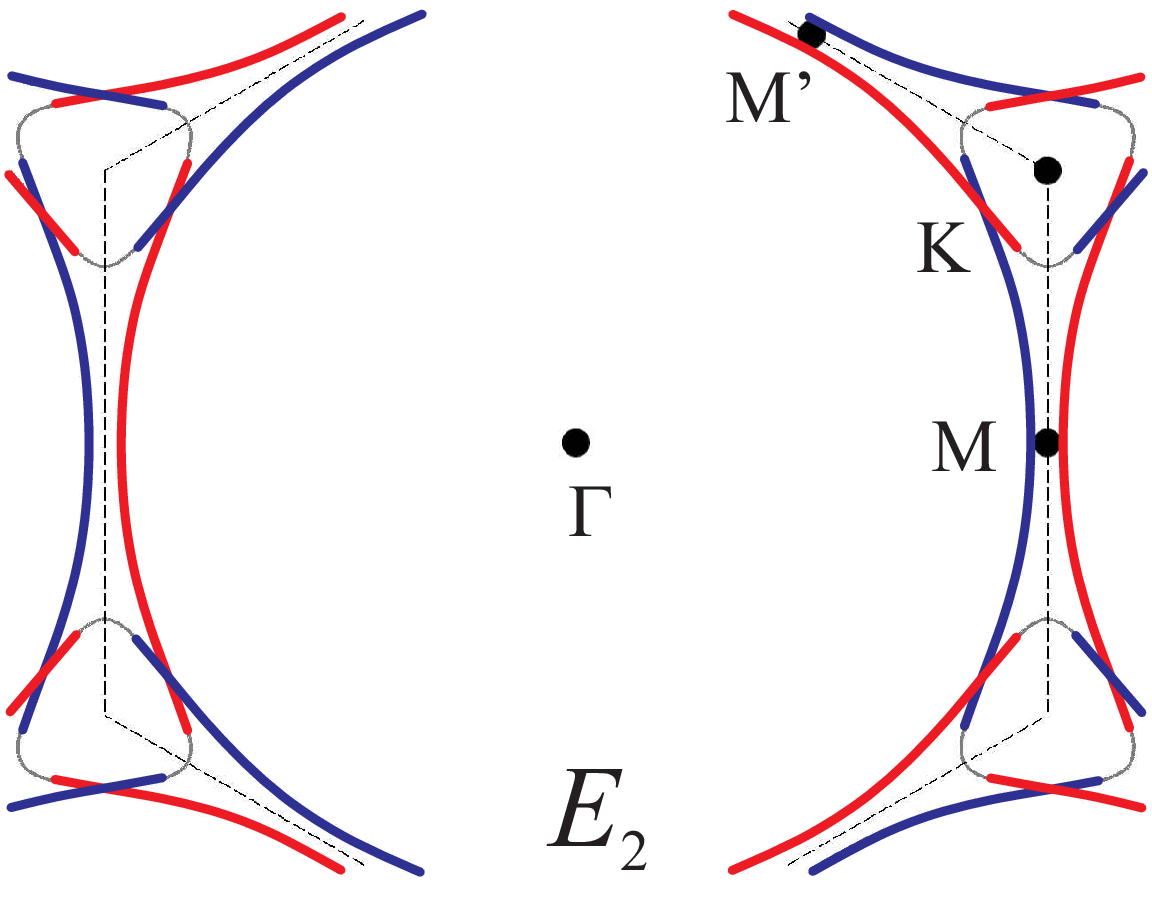}
	\caption{\label{f5} The Fermi surface in the hexagonal magnetic Brillouin zone for the two positions of the Fermi level in Fig.~\ref{f2}: $E_1$ and $E_2$. The arrows show umklapp scattering vectors.}
\end{figure}

In the 2D system a similar situation also appears as one can see in Fig.~\ref{f2}. The shape of dispersion curves along the M'-$\Gamma$-M path are qualitatively the same as for the 1D model. The Fermi surface for two positions of Fermi level ($E_1$ and $E_2$) are shown in Fig.~\ref{f5}. It should be mentioned that the Fermi surface with pockets around the K points (the left panel of Fig.~\ref{f5}) is similar to that was observed experimentally in PdCrO$_2$ \cite{Hicks2}. A spin texture of the Fermi surface emerges in the both cases. It strongly influences transport properties \cite{Kudasov,KudasovPRB}. For example, umklapp scattering under an electron-phonon interaction defines a resistivity of pure crystalline substances at low temperatures \cite{ziman}. Since the electron-phonon interaction is spin conserving, it is strongly suppressed by the spin texture, as shown in Fig.~\ref{f5}. This leads to the highly conductive state as discussed earlier \cite{Kudasov}. 

\begin{figure}
	\includegraphics[width=0.45\textwidth]{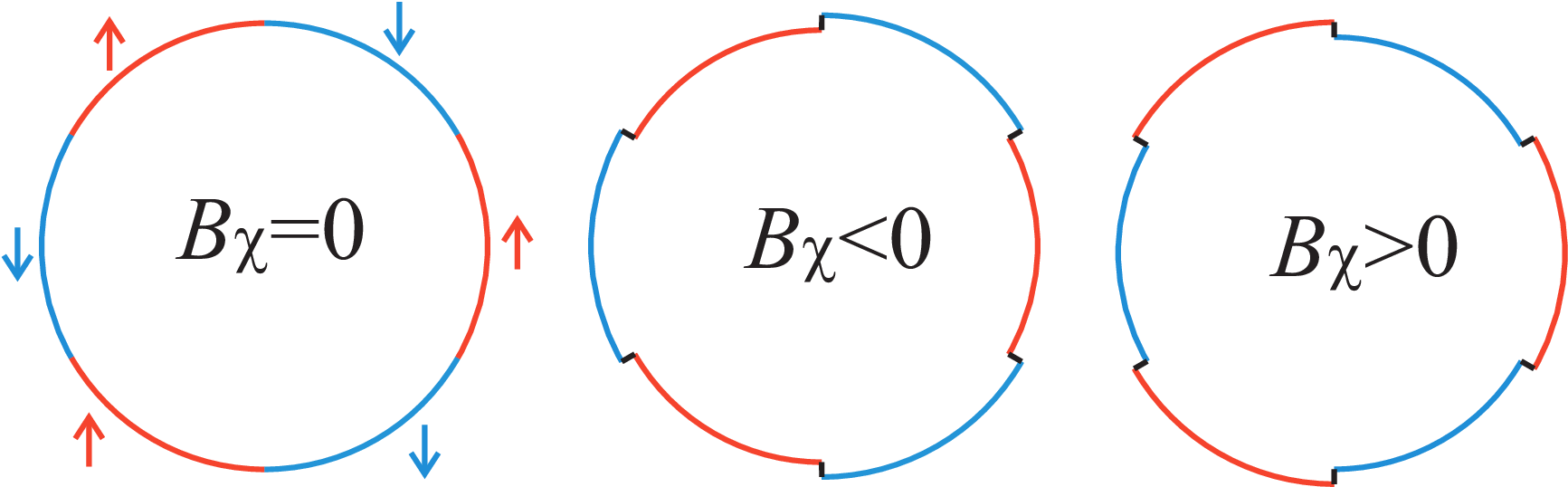}
	\caption{\label{f6} The quasi-isotropic 2D model of the Fermi surface with spin texture.}
\end{figure}

To discuss specific transport phenomena, a simple model of the Fermi surface can be considered (see Fig.~\ref{f6}).  The dispersion is assumed to be isotropic and parabolic. Then, in the absence of magnetic field ($B_{\chi} =0$) the 2D Fermi surface is a circle. It is assumed to be spin-textured, i.e. six equal arcs are alternatively spin polarized along the chirality vector as shown in the left panel of Fig.~\ref{f6}. The component of magnetic field along the chirality vector raises or lowers the dispersion surface depending on the spin direction. That is why, the Fermi surface becomes asymmetric in this case (Fig.~\ref{f6}, $B_{\chi} \ne 0$). The arcs are assumed to be fully spin-polarized, and transition regions between the arcs are neglected in the model. A difference between the Fermi wave vectors for opposite spins ($\Delta k_F= k_{F\uparrow} - k_{F\downarrow}$) at $B_{\chi} \ne 0$ is assumed to be small ($\Delta k_F \ll k_F$), and for the parabolic dispersion  
\begin{eqnarray}
	\Delta k_F = \frac{2\mu_B B_{\chi}}{\hbar v_F}
	\label{dkF}
\end{eqnarray}
where $v_F$ is the Fermi velocity at $B_{\chi} =0$.

\subsection{Kinetic equation}

Let us assume that a uniform  external magnetic field $\mathbf{B}$ is parallel to $\mathbf{n}_{\chi}$, i.e. $B_{\chi}$ is its only component, and to an external electric field $\mathbf{E}$. All of them are oriented along the crystallographic $a$-axis, which corresponds to the $\Gamma$-M direction of the magnetic Brillouin zone. In the case of $\mathbf{B} \Vert \mathbf{E}$ and isotropic parabolic dispersion, the kinetic equation is equivalent to that without magnetic field \cite{Blatt}:
\begin{eqnarray}
	\frac{e}{\hbar} \mathbf{E} \mathbf{{\nabla}}_\mathbf{k} f = -\frac{\delta f}{\tau}
	\label{df}
\end{eqnarray}
where $\mathbf{k}$ is the wave vector, $\tau$ is the relaxation time, and $f \equiv f(\mathbf{k},\mathbf{r})$ is the distribution function, which is defined so that the average number of electrons in the differential of volume $\mathrm{d} \mathbf{k} \mathrm{d} \mathbf{r}$ is $(2\pi)^{-2} f(\mathbf{k},\mathbf{r}) \mathrm{d} \mathbf{k} \mathrm{d} \mathbf{r}$. Then, the only effect of the magnetic field is the distortion of the textured Fermi surface due to Zeeman splitting as shown in Fig.~\ref{f6}.

The right side of Eq.~(\ref{df}) can be considered as a correction to the equilibrium distribution. Then, the first-order correction to the equilibrium distribution $f_0$ takes its usual form \cite{Blatt}: 
\begin{eqnarray}
	f_1 = e \tau E_x v_x \frac{\mathrm{d} f_0}{\mathrm{d} \varepsilon} 
	\label{df1}
\end{eqnarray}
Here $E_x$ is the only component of the electric field strength, the particle energy has the standard form $\varepsilon = \hbar^2 k^2 /(2m)$, and the particle velocity is $\mathbf{v}=\hbar^{-1} \mathbf{{\nabla}}_\mathbf{k} \varepsilon$. 

The current density in the 2D model is defined by the following expression 
\begin{eqnarray}
	j_x = \frac{e}{4\pi^2} \int  v_x f \mathrm{d} \mathbf{k}.
	\label{j0}
\end{eqnarray}
The first-order correction leads to a linear response, i.e. ordinary conductivity with a minor correction due to the Fermi surface distortion.
  
\subsection{Nonreciprocal electronic transport}

An equilibrium distribution function in a normal metal is centrally symmetrical, i.e. $f_0(\mathbf{k}) = f_0(\mathbf{-k})$. Then the first-order correction is antisymmetrical $f_1(\mathbf{k}) = - f_1(\mathbf{-k})$, and the second-order one should be again centrally symmetrical $f_2(\mathbf{k}) = f_2(\mathbf{-k})$. That is why, the second-order correction to current disappears, and the electron transport turns out reciprocal. However, this statement is obviously invalid in case of the spin-textured Fermi surface at $B_\chi \neq 0$.

The second-order correction to the distribution function is obtain by substitution $f_1 \rightarrow f$ to the left side of Eq.~(\ref{df}):  
\begin{eqnarray}
	f_2 = (e \tau E_x)^2 \left[ \frac{\mathrm{d}^2 f_0}{\mathrm{d} \varepsilon^2} v_x ^2 + \frac{1}{m}\frac{\mathrm{d} f_0}{\mathrm{d} \varepsilon} \right].
	\label{df2}
\end{eqnarray}
The second-order correction to current density $j_2$ is got by substitution of Eq.~(\ref{df2}) to Eq.~(\ref{j0}). The second term of Eq.~(\ref{df2}) gives the zero contribution to the current because it is isotropic. Then, the correction takes the form   
\begin{eqnarray}
	j_2 = \frac{e (e \tau E_x)^2}{\sqrt{2}\pi^2 m^{1/2} \hbar^2} \int \int \frac{\mathrm{d}^2 f_0}{\mathrm{d} \varepsilon^2} \varepsilon^{3/2}\mathrm{d} \varepsilon \cos^3(\phi) \mathrm{d} \phi.
	\label{j2}
\end{eqnarray}
In the low-temperature limit, $T/\varepsilon_F \ll 1$, the equilibrium distribution function goes to the $\Theta$-function. A finite temperature leads to a small correction to $j_2$, which is itself a second order correction, that is, it can be neglected. Then, applying rules of generalized functions differentiation \cite{Vladimirov} we obtain 
\begin{eqnarray}
	j_2 = - \frac{3 e (e \tau E_x)^2}{8 \pi^2 \hbar^2} \Delta v_F.
	\label{j21}
\end{eqnarray}

One can represent conductivity as a sum of linear and nonlinear parts:
\begin{eqnarray}
	\sigma = \sigma_0 + \sigma_2 E 
	\label{sig}
\end{eqnarray}
where $\sigma_2$ takes the final form by means of Eq.(\ref{dkF}):
\begin{eqnarray}
	\sigma_2 = - \frac{3 e^3 \tau^2 \mu_B}{4 \pi^2 \hbar^2 m v_F} B_{\chi}
	\label{s2}
\end{eqnarray}

Since the experimental data is expressed in Ref.~\cite{Akaike} in terms of resistivity
\begin{eqnarray}
	\rho = \rho_0 + \rho_2 j, 
	\label{sig rho}
\end{eqnarray}
it is convenient to reformulate Eq.(\ref{s2}) as follows
\begin{eqnarray}
	\rho_2 \approx - \frac{\sigma_2}{\sigma_0^3} \propto \frac{B_{\chi}}{\tau}.
	\label{r2}
\end{eqnarray}
The second terms in Eq.~(\ref{sig}) and Eq.~(\ref{sig rho}) are assumed to be small as compared to the first ones.

A temperature dependence of the non-linear part of conductivity is determined by the relaxation time ($\tau$). That is why, $\rho_2 (T)$ occurs nonmonotonic. It is small at very low temperatures, then grows with the temperature due to   the decrease of the relaxation time, and disappears at the Neel point. It is also easy to see that the ratio $\rho_2/\rho_0$ is temperature-independent. These conclusions are in a qualitative agreement with the experimental results \cite{Akaike}.

\begin{figure}
	\includegraphics[width=0.35\textwidth]{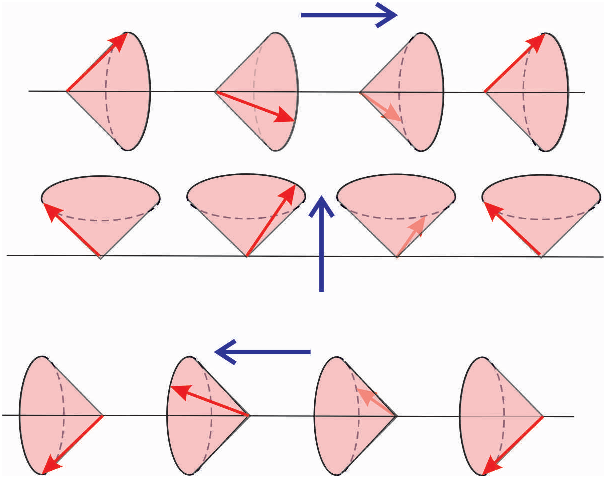}
	\caption{\label{f7} A transformation of helical structure under rotating magnetic field.}
\end{figure}

An interpretation of the experimental data on nonreciprocal transport is complicated at low magnetic field by magnetic anisotropy. On the other hand, if the external magnetic field is above the spin-flop transition ($B_{sf}$), the chirality vector approximately follows by the magnetic field. For PdCrO$_2$, $B_{sf}$  is estimated to be about 6.5~T \cite{Hicks2}. As a consequence, a specific effect in high magnetic fields appears.
$\rho_2 (B)$ is an odd function for a certain direction. However, a rotation of the magnetic field about a perpendicular axis by angle $\pi$ leads to the same value of $\rho_2$ because it alternates the chirality together with the external magnetic field as schematically shown in Fig.~\ref{f7}.

There is another difficulty for a quantitative analysis of the experimental results of Ref.\cite{Akaike}. Sizes of sample cross-sectional area ($1 \times 1.85$~$\mu \mathrm{m}^2$) were smaller than an estimation of mean free path at low temperatures ($l_f \approx 4$~$\mu \mathrm{m}$ at $T=4$~K \cite{Mackenzie}). This leads to anisotropy of the relaxation time \cite{Fuchs,ziman}. It can enhances the nonreciprocal effect. Moreover, since the mean free path decreases with growth of temperature there may exist a transition to the isotropic regime at a higher temperature.  

\section{Discussion and conclusion}

The two theorems proved above show that Kramers-like degeneracy exists in a helical magnetic field. In contrast to theorem 1, we did not extended theorem 2 to the interaction of particle charge and magnetic field because in the latter case a strong additional assumption would have to be required. That is, Hamiltonian $\hat{H}_0$ should be invariant not only with respect to spin rotations but also to spatial ones.

The Kramers-like symmetry of electron dispersion in helical magnetic (or effective) field together with generalized Bloch theorem and periodic boundary conditions create the non-trivial topology of band structure. It is not related to Berry phase but stems from multisheet structure of dispersion curves or surfaces. In the present article we discussed 1D and 2D systems. However the theory can be extended to 3D ones because the theorems proved and topological arguments remain valid in this case as well.

As a result of the nontrivial topology an isolated non-periodic dispersion curve may appear, and an unusual transport phenomena can be observed. That is why, we called this case a strong topological metal. While few bands crossing the Fermi level, an interband scattering suppresses the highly conductive state. The nonreciprocity is also smeared.

The 1D and 2D models discussed in the previous sections have much in common with the unconventional $p$-wave antiferromagnet \cite{Hellenes,Brekke}. For instance, it falls into the scope of theorem 1.The 1D electron dispersion of the $p$-wave antiferromagnet possesses the Kramers-like symmetry with the exception of special points at $k=\pm \pi$ \cite{Brekke}. If the Fermi level is close to the bottom in the band structure of Fig.~\ref{f2}, two separate Fermi surfaces for spin-up and spin-down electrons appear. Because of the hexagonal symmetry of the model their shape corresponds to a $f$-wave antiferromagnet in terms of Ref.~\cite{Brekke}. However, if the Fermi level fits the strong topological metal, e.g. at $E_1$ in Fig.~\ref{f2}, there exists a single spin-textured Fermi surface (Fig.~\ref{f2}). This distinguishes the strong topological metal state from $p$-wave antiferromagnets and altermagnets and makes a novel class of unconventional magnets \cite{Smejkal,Hellenes,Brekke}.

The spin texture of the Fermi surface gives rise to anomalous high-conductive state in 2D systems \cite{Kudasov} and  nonreciprocity. This behavior is similar to that of topological surface and edge states in topological insulators \cite{Bansil}. However, in the present work we dealt with bulk states. The previous experimental investigation of nonreciprocity in PdCrO$_2$ \cite{Akaike} was motivated by phenomenological arguments, i.e. Onsager's relations for electronic transport in chiral systems \cite{Rikken}. Here, we provided its microscopic description.

In the present research, we considered PdCrO$_2$ as an example of 2D helical system. That is, we assumed that hexagonal palladium layers are under the 120$^0$ effective field of chromium layers with the non-zero net chirality.
This point of view is supported by experimental observations of non-reciprocal electronic transport \cite{Akaike} and UAHE \cite{Takatsu3}. However, the thorough study of the magnetic structure of PdCrO$_2$ performed by neutron scattering \cite{Takatsu2} revealed an alternated chirality of chromium layers. That is why, a pair of adjacent Cr layers should produce the total effective field with the zero net chirality. This contradiction is partially resolved by the fact that a correlation length of the magnetic structure of PdCrO$_2$ is rather small along the $c$-axis \cite{Billington} and the nature of the magnetic order is quasi-2D in this substance. Therefore, there is a sizable fraction of palladium layers, which are under the effective field with non-zero chirality at domain boundaries.

It is difficult to meet all the conditions of strong topological metal, i.e. layers with high intrinsic conductivity, helical spinor effective field, and a single band at the Fermi level. The only candidate discussed above is PdCrO$_2$. That is why, van der Waals laminated systems may provide an opportunity to create substances with the topologically nontrivial band structures because they allow alternating conductive and magnetic layers \cite{Chang,Lei}. The example of PdCrO$_2$ shows that the topological metals are the way to materials with extremely high conductivity.  

\begin{acknowledgments}
	The author thanks Prof.~A.~A.~Fraerman and Prof.~V.~M.~Uzdin for fruitful discussions. 
	The work was supported by National Center for Physics and Mathematics (Project \#7 "Investigations in high and ultrahigh magnetic fields").
\end{acknowledgments}

\section*{Appendix A: Fourier coefficients of helical spinor field }

Let us define a general form of Fourier coefficients in Eq.~(\ref{NFE}) for a special case of helical spinor field, that is, 
\begin{eqnarray}
	 \hat{\mathbf{t}} \hat{\mathbf{r}}_{\alpha} \mathbf{h}(\mathbf{r}) \hat{\mathbf{ \boldsymbol\sigma}} \hat{\mathbf{r}}_{\alpha}^{-1} \hat{\mathbf{t}}^{-1}  =\mathbf{h}(\mathbf{r}) \hat{\mathbf{ \boldsymbol\sigma}},
	  \label{helical}
\end{eqnarray}
where $\alpha=2\pi/n$, $\hat{\mathbf{t}}=\hat{\mathbf{T}}/n$, $n>2$ is an integer number. $\mathbf{T}$ is a primitive vector of the magnetic lattice.

The Fourier coefficients has the following form \cite{Ashcroft}: 
\begin{eqnarray}
	\hat{U}_{\mathbf{K}} = \frac{1}{V} \int \exp(-i \mathbf{K} \mathbf{r}) \mathbf{h}\left( \mathbf{r} 
	\right) 	\hat{\boldsymbol{\sigma}} \mathrm{d}\mathbf{r} \label{UK}
\end{eqnarray}
where $\mathbf{K}$ is a primitive vector of reciprocal lattice, i.e. $\mathbf{K} \mathbf{T} = 2\pi$.

For the sake of simplicity, we also assume that $\mathbf{h}(\mathbf{r})$ lies in the $xy$ spin plane. Then, the Fourier coefficient can be written down as  
\begin{eqnarray}
	\hat{U}_{\mathbf{K}} = \left(
	\begin{array}{cc}
		0 & a\\
		b & 0
	\end{array}
	\right)
	\label{U}
\end{eqnarray}
where $a$ and $b$ are complex coefficients. The translation of the effective field to vector $\mathbf{t}$ gives an additional factor $\exp(i \mathbf{K} \mathbf{t})=\exp(i \alpha)$ in Eq.~(\ref{UK}). After the rotation of the spinor about $z$ axis by angle \cite{Landau} and taking into account Eq.~(\ref{helical})
we obtain
\begin{eqnarray}
	\hat{U}^{\prime}_{\mathbf{K}} = e^{i \alpha} \left(
	\begin{array}{cc}
		0 & a e^{-i \alpha} \\
		b e^{i \alpha}  & 0
	\end{array}
	\right) =  \left(
	\begin{array}{cc}
		0 & a \\
		b e^{2 i \alpha} & 0
	\end{array} 
	\right). 
	\label{U} 
\end{eqnarray}
The solution of equation $\hat{U}^{\prime}_{\mathbf{K}} = \hat{U}_{\mathbf{K}}$ is $b=0$ and $a$ is arbitrary. That is why, the general form of the Fourier coefficients is
\begin{eqnarray}
	\hat{U}_{\mathbf{K}} = a \left(\hat{\mathbf{\sigma}}_{x} \pm i \hat{\mathbf{\sigma}}_{y}\right).
	\label{U1}
\end{eqnarray}  
Here the sign corresponds to the left or right chirality.

The coefficients of this type were obtained for 1D and 2D models by direct calculations in Refs.~(\cite{Kudasov,KudasovFTT,KudasovPRB}).

It should be noted that operators Eq.~(\ref{U1}) are abnormal \cite{Kudasov}, that is, 
\begin{eqnarray}
	\hat{U}_{\mathbf{K}} \hat{U}_{\mathbf{K}}^{\dagger} \ne \hat{U}_{\mathbf{K}}^{\dagger} \hat{U}_{\mathbf{K}}.
	\label{U2}
\end{eqnarray}  
This leads to an important consequence. The solutions of Eq.~(\ref{NFE}) become noncentro-symmetrical since
\begin{eqnarray}
	\hat{U}_{\mathbf{K}} \hat{U}_{\mathbf{K}}^{\dagger} \ne \hat{U}_{\mathbf{-K}} \hat{U}_{\mathbf{-K}}^{\dagger}.
	\label{U2}
\end{eqnarray}  

\section*{Appendix B: chirality vector  for helical effective field}

We define the chirality vector in three-sublattice tight-binding  models (1D and 2D) as follows:
\begin{eqnarray}
	\mathbf{ \boldsymbol\chi} = \mathbf{h}_1 \times \mathbf{h}_2 + \mathbf{h}_2 \times \mathbf{h}_3 + \mathbf{h}_3 \times \mathbf{h}_1. \label{chir}
\end{eqnarray}
Here $\mathbf{h}_i$ is the effective field on sites of the i-th sublattice.

In case of a continuous 1D or 2D model, where $\mathbf{h}(\mathbf{r})$ is a continuous periodic function in the magnetic unit cell, one can generalize Eq.~(\ref{chir}). Let us consider a system satisfying the conditions of theorem 2. Under the shift by vector $\mathbf{t}=\mathbf{T}/3$ the effective field rotates about the $z$ axis by angle $\alpha=2\pi/3$. $\mathbf{T}$ is a primitive vector of the magnetic lattice. Then, the chirality vector can be written down as
\begin{eqnarray}
	\mathbf{ \boldsymbol\chi} = \frac{1}{V_c}\int  \mathbf{h}(\mathbf{r}) \times \mathbf{h}(\mathbf{r}+\mathbf{t}) \mathrm{d}\mathbf{r}  \label{chir1}
\end{eqnarray}
Here, the integration is performed over the magnetic Brillouin zone. 
A specific direction (through vector $\mathbf{t}$) enters into this definition. It plays the same role as the order of sublattices numbering in discrete models (Eq.~(\ref{chir})).
This definition of vector chirality is not universal but useful in our particular case.

It is also useful to introduce a unit chirality vector:
\begin{eqnarray}
	\mathbf{n}_{\chi} = \frac{\mathbf{ \boldsymbol\chi}}{|\mathbf{ \boldsymbol\chi}|}.
	\label{chir2}
\end{eqnarray}

\bibliography{topmet}

\begin{thebibliography}{56}
\bibitem [{not()}]{note}%
  \BibitemOpen
  \href@noop {} {}\bibinfo {note} {To be precise, there is no ambient isotopy
  \cite{Kalajdzievski} from an ordinary dispersion surface to multisheet
  one.}\BibitemShut {Stop}%
\end{thebibliography}

\end{document}